\documentclass[10pt, a4paper,twocolumn]{article}
\usepackage[utf8]{inputenc}
\usepackage{cite}
\usepackage{amsmath,amssymb,amsfonts}
\usepackage{algorithmic}
\usepackage{algorithm}
\usepackage{graphicx}
\usepackage{textcomp}
\usepackage{booktabs}
\usepackage{subcaption}
\usepackage{xcolor}
\usepackage{hyperref}
\usepackage{geometry}
\title{Preference-Based Privacy Trading}

\newtheorem{theorem}{Theorem}[section]

\usepackage{wrapfig}
\usepackage{graphicx}
\usepackage{amsmath}
\usepackage{amsfonts}
\usepackage{amssymb}
\usepackage{color}
\usepackage{tabularx}
\usepackage{multirow}
\usepackage{breqn}
\usepackage{tabu}
\usepackage{float}
\usepackage{caption} 
\usepackage{algorithmic}

\usepackage{fancyhdr}
\usepackage{tikzpagenodes}
\begin{document}

\title{Aggregate Cyber-Risk Management in the IoT Age: 
\\ \emph{{Cautionary Statistics for (Re)Insurers and Likes}}
}

\author{
\uppercase{Ranjan Pal},
\uppercase{Ziyuan Huang},
\uppercase{Xinlong Yin},
\uppercase{Sergey Lototsky},\\
\uppercase{Swades De},
\uppercase{Sasu Tarkoma},
\uppercase{Mingyan Liu},
\uppercase{Jon Crowcroft},
\uppercase{Nishanth Sastry}
}
\date{}
\maketitle

\begin{abstract}
IoT-driven smart societies are modern service-networked ecosystems, whose proper functioning is hugely based on the success of supply chain relationships. 
Robust security is still a big challenge in such ecosystems, catalyzed primarily by naive cyber-security practices (e.g., setting default IoT device passwords) on behalf of the ecosystem managers, i.e., users and organizations. 
This has recently led to some catastrophic malware-driven DDoS and ransomware attacks (e.g., the \emph{Mirai} and \emph{WannaCry} attacks). 
Consequently, markets for commercial third party cyber-risk management services (e.g., cyber-insurance) are steadily but sluggishly gaining traction with the rapid increase of IoT deployment in society, and provides a channel for ecosystem managers to transfer residual cyber-risk post attack events. Current empirical studies have shown that such residual cyber-risks affecting smart societies are often \emph{heavy-tailed} in nature and exhibit \emph{tail dependencies}. 
This is both, a major concern for a profit-minded cyber-risk management firm that might normally need to cover multiple such dependent cyber-risks from different sectors (e.g., manufacturing, energy) in a service-networked ecosystem, and a good intuition behind the sluggish market growth of cyber-risk management products. 
In this paper, we provide (i) a rigorous general theory to elicit conditions on (tail-dependent) heavy-tailed cyber-risk distributions under which a risk management firm might find it (non)sustainable to provide aggregate cyber-risk coverage services for smart societies, and (ii) a real-data driven numerical study to validate claims made in theory assuming \emph{boundedly rational} cyber-risk managers, alongside providing ideas to boost markets that aggregate dependent cyber-risks with heavy-tails. \emph{To the best of our knowledge, this is the only complete general theory till date on the feasibility of aggregate cyber-risk management.}

\end{abstract}

\section*{keywords}
aggregate cyber-risk, heavy-tail, tail-dependency


\section{Introduction}
 
IoT-driven smart cities are examples of service networked ecosystems that are popularly on the rise around the globe, with major cities like Singapore, Dubai, Barcelona, and Amsterdam being working examples. The proper functioning of such cities is hugely based on the success of supply chain relationships from diverse sectors such as automobiles, electronics, energy, finance, aerospace, etc. In the IoT age, these relationships are often realized via large scale systemic network linkages (see Figure 1.1. in \cite{coburn2018solving}) that operate via the interplay of IoT hardware (e.g., sensors, actuators, cameras), application software (e.g., Oracle for DBMS support, cloud service software), and IoT firmware.

Currently, robust IoT security is a challenge \cite{gilchrist2017iot} with a significant fraction of users controlling IoT systems being naive about effective cyber-security practices (e.g., the use of non-default device passwords, periodic patch updates). Consequently a cyber-attack exploiting a software vulnerability can have a catastrophic cascading service disruption effect that could amount to losses in billions of dollars across various service sectors. Recent examples of such cyber-attacks include the \emph{Mirai} DDoS (2016), \emph{NotPetya} ransomware (2017), and \emph{WannaCry} ransomware (2017) attacks, which wrecked havoc among firms in various industries across the globe, resulting in huge financial losses due to service interruption (see \cite{coburn2018solving} for more examples). As a result of such large losses, a certain section of society overall could be negatively impacted and experience psychological depression and affected lifestyles. 

As instruments to cover cyber-losses in society, markets for commercial third-party services (e.g., cyber-insurance) are steadily but sluggishly gaining traction with the rapid increase of societal IoT deployment, and provides a channel for members (individuals and organizations) to transfer residual cyber-risk post cyber-attack events. The primary benefits of commercial cyber-loss management services have been recently cited in detail by the authors in Biener et.al. \cite{biener2015insurability}, and include (i) indemnification of loss events, (ii) helping corporations estimate cost of cyber-risk, and (iii) improve cyber-security \cite{pal2010analyzing}\cite{pal2014will}\cite{pal2018improving}\cite{pal2019robust}.  The steady rise in market requirement for such services primarily arises from a combination of (a) the naivety of user security practices, (b) the non fool-proof nature of technical security solutions to remove cyber-risk \cite{anderson2009information}, (c) higher board level concerns in organizations post notable cyber-breach incidents (e.g., Sony, Target, WannaCry) and their negative effect on stock prices \cite{shetty2018reducing} \cite{gatzlaff2010effect}, and (d) the growing perception of cyber-risk in the digital society \cite{pooser2018growth}.

Despite the promised potential for commercial cyber-risk management services, the markets have been too sluggish for our liking. The yearly estimates of cyber-loss approximately amount to USD 600 billion globally (1\% of US GDP) \cite{coburn2018solving}, whereas the cumulative global public and private sector spendings on cyber-security amount only to USD 174 billion \cite{wang2019integrated}. In addition,  the total yearly market for cyber-insurance services - the most popular form of commercial third party commercial cyber-risk management offerings, approximates to a paltry USD 6 billion globally \cite{wang2019integrated}, compared to the amount of net cyber-loss. The primary reasons for such a low (but increasing) market penetration are (a) misunderstanding and lack of coverage awareness by the demand side (users and organizations) \cite{wang2019integrated}, (b) unavailability of quality plus quantity data on cyber-risks and demand side cyber-hygiene behavior, that contribute to policy pricing nuances \cite{romanosky2019content} \cite{franke2017cyber} \cite{wang2019integrated}, and (c) the empirical evidence of certain cyber-risk distributions being heavy-tailed and tail-dependent \cite{biener2015insurability} \cite{eling2020extreme} \cite{xu2018modeling} \cite{maillart2010heavy}, that makes profit-minded risk-averse cyber-insurers go low on confidence to expand coverage markets, where coverage is on an aggregate sum of such heavy-tailed cyber-risks. 

\subsection{Research Motivation}
{It is obvious that the ushering pervasive IoT age with 100s of IoT devices per home/organization will bring forth the need for businesses and homes to increasingly buy coverage CRM solutions like cyber-insurance. This is simply because the cyber-attack space will be broad enough in the digital terrain for humans to always prevent being security-hacked by smart adversaries. As a result, any coverage CRM solution provider will face aggregate cyber-risks from its clients. The idea of spreading aggregate cyber-risk among multiple risk managers (e.g., cyber (re)insurers) is gaining traction \cite{coburn2018solving} \cite{kessler2014re} \cite{li2018scor} for IoT-driven smart society settings whereby insurers covering aggregate cyber-risk of organizations in a given sector (e.g., manufacturing) wish to spread that risk among insurers of firms that are higher up in the supply chain (e.g., energy companies). However (a) there is \textbf{no formal analysis} on the effectiveness of this idea for \emph{general individual cyber-risk distributions}, and (b) there may be significant differences in the cyber and non-cyber re-insurance settings - benefits of non-systemic outcomes in the latter (as qualitatively stated in \cite{kessler2014re}) may not apply to the former (see Section IV for more details). Consequently, without a formal analysis, aggregate cyber-risk managers may not have the confidence to scale their service markets \cite{welburn2019systemic}. \emph{Our main goal in this paper is to devise a foundational methodology that analyzes the effect of individual heavy-tailed and tail-dependent cyber-risks on the effectiveness of aggregate cyber-risk management markets.}}

\subsection{Research Contributions}
We make the following research contributions in this paper.
\begin{enumerate}
    \item We prove that spreading \emph{catastrophic} heavy-tailed cyber-risks that are identical and independently distributed (i.i.d.), i.e., not tail-dependent, \emph{is not} an effective practice for aggregate cyber-risk managers. However, spreading i.i.d. heavy-tailed cyber-risks that are \emph{not catastrophic} is an effective practice for aggregate cyber-risk managers. While this latter point has long been believed and empirically validated in the cyber-insurance research literature, the former point is a surprising new facet that we unravel in this paper via theory (see Section II).
    \item We prove that spreading \emph{catastrophic} and \emph{curtailed} heavy-tailed cyber-risks that are (non) identical and independently distributed (i.i.d.), i.e., not tail-dependent, \emph{is not} an effective practice for aggregate cyber-risk managers. (see Section III).
    \item We show that spreading \emph{catastrophic and tail-dependent} heavy-tailed cyber-risks \emph{is not} an effective practice for aggregate cyber-risk managers. Though this result has been empirically established in the past for some heavy-tailed distributions (and also somewhat intuitive from the results of Section II), \emph{there exists no formal proof for general heavy-tailed cyber-risk distributions, leave alone catastrophic heavy-tailed distributions} (see Section IV). 
    \item We experimentally validate our theory using a real-world Privacy Rights Clearinghouse cyber-breach data set (see Section V). 
\end{enumerate}
Our proposed research, based on ideas in \cite{ibragimov2015heavy} presents a foundational methodology to analyze the effectiveness of spreading catastrophic heavy-tailed and tail-dependent cyber-risks. \emph{To the best of our knowledge, this is the only complete general theory till date on the feasibility of aggregate cyber-risk management, and is {
invariant of specific threat models that eventually induce cyber-risk distributions}.}
Though the empirical occurrence of catastrophic cyber-risks is uncommon, it is a matter of time we start encountering them relatively more frequently in the IoT age (see Chapters 1.2, 1.3 in \cite{coburn2018solving}). \emph{A basic primer of important statistical and econometric concepts used in the paper is provided in {\href{https://drive.google.com/file/d/190WaqylxZCTpMCchynYaJ_I17zrBIMfr/view?usp=sharing}{ \underbar{online Appendix A}}}}, and a table of important notations in the paper is presented in Table I.

\subsection{Contribution Impact on Society and Technology}
{Our research contributions stated thus far are primarily targeted towards the advancement in the economics and econometrics of cyber-risk management in the IoT age through the solution of open research issues - the main focus of our research. However, each of these contributions have a \emph{direct impact} on IoT security improvement, and its consequent positive impact on society. 

To start with, according to data sources, the global number of connected devices has already reached 22 billion at the end of 2018 - more than half of which belong to enterprise IoT \cite{iotconnected1}, and will grow to 29 billion by 2022 \cite{iotconnected2}. 
Moreover, worldwide spending on IoT is projected to reach a significant 1.2 trillion USD by 2022 with the number of Internet-connected devices being projected to reach a whopping 125 billion by 2030 \cite{markit2017internet}. 
\emph{A thing common to nearly all IoT devices is the poor cyber-hygiene associated with their use (e.g., default passwords) - a primary reason being the scale of such devices in operation and the disproportionate human effort (that is likely to continue) needed to strengthen basic security in such devices \cite{coburn2018solving}.} 
This increasingly becoming common knowledge would push organizations and individual households to consider investing in third-party cyber-risk management (CRM) solutions as a necessary risk management step in the upcoming pervasive IoT age.

\textbf{Contribution \#1} states that cyber-risk ``buyers'' (i.e., the CRM firms) need to develop regulated pricing policies for their CRM solutions. These solutions will enable end-users to voluntarily (incentive compatibly) ``look after" to a considerable degree, the security hygiene (and hence cyber-risk exposure) of IoT devices under their control. Consequently, such steps will prevent each end-user (individual household or organization) to be a source of a cyber-risk distribution that is heavy-tailed, i.e., catastrophic. 
This will allow CRM solution markets to scale and flourish, and improve cyber-security in society. 
\textbf{Contribution \#2} reflects the same things for the CRM solution buyers as that from Contribution \#1, but additionally warns the `risk-buyer' side to put increasing focus on pricing policies that prevent IoT-controlled sources (organizations or individual households) to be a root of catastrophic cyber-risk distributions. The increased focus needed due to the fact that statistical curtailement of such cyber-risks (unlike that in Contribution \#1) will also not allow CRM markets to scale and flourish - thereby having a negative effect on society as a whole. 
\textbf{Contribution \#3} reflects similar learnings for both the CRM solution provider and the buyer sides, as that from Contributions\#'s 1 and 2. 
\textbf{Contribution \#4} clearly states that when CRM solution providers suffer from practical and subjective behavioral biases in appropriately assessing cyber-risk extent \cite{coburn2018solving}, it should not aggregate cyber-risk of catastrophic nature - thereby implying, similar to that in Contribution \#'s 1-3, that solution pricing policies should be designed in a way so as to incentivize CRM solution buyers to invest enough efforts in cyber-security so as not to be a source for catastrophic cyber-risks. 
Finally, while appropriate CRM pricing policies might `nudge' the demand side to improve their cyber-hygiene, all the contributions together indicate the important role of regulators (e.g., the government) to regulate the enforcement of improved security strength in factory settings of IoT devices during/post manufacturing. 
This will mitigate (a) the negative effect of human ``laziness'' towards improving cyber-hygiene, and (b) the chances of society dealing with catastrophic risks.}

\section{(Catastrophic) IID Cyber-Risk Aggregation} 
\label{sec:analysis}
One of the key features of risk management (CRM) (e.g., via insurance) in general as a business model is its ability to pool different types of risks, thereby reducing an underwriter's overall risk exposure.  This is particularly true for a reinsurer (not necessarily a cyber re-insurer) , who is in a position to significantly diversify its risks, by selling reinsurance contracts to very different front-line insurers who specialize in different sectors (e.g., retail, pharmaceutical, manufacturing, etc.), primarily independent of one another. This means that a reinsurer typically takes on or \emph{aggregates} a fraction of many different risks that are most likely to be independent of one another. However, this independence property may not hold true of some cyber-risks. \emph{In Section II \& III, we make a simplistic assumption that cyber-risks aggregated by a aggregate cyber-risk manager are independent, and leave the analysis of tail-dependent cyber-risks for Section IV.} Specifically, in this paper we will often consider the average of $n$ (dependent or independent) cyber-risks $X_1, \cdots, X_n$ arising from different IoT-driven organizations in a smart society, given by $Z_{\underline{w}} = \frac{1}{n}\sum_{i=1}^{n}X_{i}$, or more generally, the weighted average given a fraction of each cyber-risk ${w}=[w_1, \cdots, w_n]$: $Z_{{w}}= \sum_{i=1}^{n} w_i X_i$. 
In what follows, in this section we will first examine, for increasing cyber-risk spread (variance), the distribution resulting from aggregating catastrophic cyber-risks, whose first and second moments are undefined. We will then generalize this result and examine the standard VaR risk measure (\emph{see {\href{https://drive.google.com/file/d/190WaqylxZCTpMCchynYaJ_I17zrBIMfr/view?usp=sharing}{ \underbar{online Appendix A}}  } for a definition and a valid rationale for using the VaR metric}) as a result of aggregating $n$ cyber-risks (catastrophic or otherwise). 

\begin{table}[htbp]
\centering 
\begin{tabular}{r p{6cm} }
\hline
Symbol & Description \\
\hline
$VaR_q(X)$ &  Value-at-Risk (VaR) of X at level q\\
$S_\alpha(\sigma, \beta, \mu)$ &   stable and heavy-tailed distribution characterized by the index of stability $\alpha$, scale parameter $\sigma$, symmetry index $\beta$, and location parameter $\mu$\\
$\mathcal{\overline{CS}}(r)$  & class of symmetric distributions that are convolutions of $S_\alpha(\sigma, 0, 0)$ distributions with $r\leq \alpha < 2$ and $\sigma > 0$\\
$\mathcal{\underline{CS}}(r)$  & class of symmetric distributions that are convolutions of $S_\alpha(\sigma, 0, 0)$ distributions with $0\leq \alpha < r$ and $\sigma > 0$\\
$\mathcal{CSLC}$ &   class of symmetric distributions that are convolutions of symmetric distributions that are either log-concave or stable with exponent $\alpha > 1$ \\  
$Z_{w}$ & aggregated risk with weights $w$ and risk portfolio $X_1,\cdots,X_n$, such that $Z_w=\sum_{i=1}^{n}w_iX_i$\\
$a$ & length of support of a probability distribution\\
\hline
\end{tabular}
\caption{Table of Notation}
\label{tab:TableOfNotation}
\end{table}
\subsection{An intuitive observation} 
To give some intuition, we begin with a simple comparison of risk spread (standard deviation) between aggregating light-tailed distributions and heavy-tailed distribution.  Consider the \emph{Normal} distribution as a representative of the former and the \emph{Levy} \cite{forbes2011statistical} and the \emph{Cauchy} distributions as representatives of the latter that are \emph{statistically stable}\cite{zolotarev1986one}; 
the latter exhibit power-law decay with cdf given by $F(-x) \approx x^{-\alpha}, x, \alpha > 0$. 
For $n$ IID normal $X_{1}, \cdots, X_{n} \sim \mathcal{N}(\mu, \sigma^{2})$, their average $\frac{1}{n}\sum_{i=1}^{n}X_{i}$ is also normally distributed with $\mathcal{N}(\mu, \frac{1}{n}\sigma^{2})$.  The implication here is that the aggregate risk has a spread (the standard deviation) that 
grows as $\sqrt{\frac{1}{n}}$ of $\sigma$ for a given $\mu$, suggesting a decrease in average risk as one spreads over an increasing number of individual risks.  Thus in this case higher diversification -- the spreading over larger pool of risks -- is desirable. 

Now consider the Levy distribution denoted by $\mathcal{L}(\mu, \sigma)$, with location parameter $\mu$, scale $\sigma$, pdf and cdf is respectively given by 
\[
    \phi(x)=\left\{
                \begin{array}{ll}
                 \sqrt{\frac{\sigma}{2\pi}}e^{\frac{-\sigma}{2(\mu - x)}}(\mu - x)^{\frac{-3}{2}} & \mathrm{if }~ x < \mu,\\
                  0 & \mathrm{if }~ x \ge \mu, 
                \end{array}
              \right.
  \]
\[
    F(x)=\left\{
                \begin{array}{ll}
                 \frac{2}{\sqrt{\pi}}\int_{0}^{\frac{-\sigma}{\sqrt{2(\mu - x)}}}e^{-t^{2}}dt & \mathrm{if }~ x < \mu,\\
                  1 & \mathrm{if }~ x \ge \mu. 
                \end{array}
              \right.
  \]
A simple algebraic manipulation will suggest that for IID $X_1, \cdots, X_n \sim \mathcal{L}(\mu, \sigma)$,
we have $\frac{1}{n}\sum_{i=1}^{n}X_{i} \sim \mathcal{L}(\mu, n\sigma)$. In other words, contrary to the normal case, the risk spread as a result of aggregating Levy distributions \emph{increases} linearly in the number of individual risks for a given $\mu$.  This suggests that risk aggregation in this case is undesirable. 

As another example, consider the Cauchy distribution denoted by $\mathcal{G}(\mu, \sigma)$, with location parameter $\mu$ and scale $\sigma$, pdf given by 
\[\phi(x) = \frac{1}{\pi\sigma}\frac{1}{1 + \left(\frac{(x-\mu)^{2}}{\sigma^{2}}\right)},\]
and the corresponding cdf given by 
\[F(x) = \frac{1}{2} + \frac{1}{\pi}\tan^{-1}\left(\frac{x - \mu}{\sigma}\right).\]
Again, standard results suggest that for IID $X_1, \cdots, X_n \sim \mathcal{G}(\mu, \sigma)$, we have $\frac{1}{n}\sum_{i=1}^{n}X_{i} \sim \mathcal{G}(\mu, \sigma)$, meaning that the spread of the aggregate risk is unchanged from the individual risk spread.  So in this case risk aggregation does not bring risk reduction benefit; it is neither desirable nor undesirable.  

The above suggests that the notion of spreading risks is sound when the underlying individual risks are light-tailed, but casts doubts on the wisdom of doing so when the underlying risks are heavy-tailed.  In the remainder of this section we formally establish this result using the VaR risk measure. 

\subsection{Aggregating IID catastrophic cyber-risks}
We first consider aggregating IID risks $X_i$ from the family $\mathcal{\underline{CS}}(1)$, which are class of distributions that are convolutions of symmetric and stable distributions with characteristic exponent $\alpha <1$ - those exhibiting an \emph{infinite} mean and variance, and representing catastrophic cyber-risks (see {\href{https://drive.google.com/file/d/190WaqylxZCTpMCchynYaJ_I17zrBIMfr/view?usp=sharing}{ \underbar{online Appendix A}}  } for details). We have the following result regarding VaR performance post cyber-risk aggregation, the proof of which is in \href{https://drive.google.com/file/d/1xHrLRwVOFtP8mWbwf_SuVm-lBmcRPlZO/view?usp=sharing}{ \underbar{online Appendix B}}  . 

\begin{theorem}
\label{the2.1}
\emph{Consider IID r.v's $X_{i} \sim \mathcal{\underline{CS}}(1), i = 1, \cdots, n$, $q \in (0, 1)$, and $n$-vector of weights ${w}, {v} \in {\mathbf{R}}^{n}_{+}$.} 
\emph{Then}
\begin{enumerate}
\item $VaR_{q}(Z_{{w}}) > VaR_{q}(Z_{{v}})$ \emph{if ${v} \prec {w}$ and ${v}$ is not a permutation of ${w}$; in other words, the function $VaR_{q}(Z_{{w}})$ is strictly Schur-concave in ${w} \in {\mathbf{R}}^{n}_{+}$.}

\item \emph{In particular}, $VaR_{q}(Z_{\bar{w}}) < VaR_{q}(Z_{w}) < VaR_{q}(Z_{\underbar{w}})$, $\forall {w} \in \mathcal{I}_{n}$ \emph{such that ${w} \ne \underbar{w}$ and ${w}$ is not a permutation of $\bar{w}$.}
\end{enumerate}
\end{theorem}

\textbf{Theorem Implications} - On a practical note, the theorem simply implies that when an aggregate cyber-risk covering agency is faced with covering independent and identical catastrophic cyber-risk distributions, the variance of the combined distribution increases with the number of piled up cyber-risks - \emph{simply a dampening signal for-profit cyber-risk managers to contribute to a sustainable aggregate loss coverage market}.

Now consider the special borderline case $\alpha = 1$ (borderline catastrophic), which corresponds to IID $X_{1}, \cdots, X_{n}$ with a symmetric Cauchy distribution $S_{1}(\sigma, 0, 0)$. 
In this case, we have for all ${w} = (w_{1},.....,w_{n}) \in \mathcal{I}_{n}$, $Z_{{w}} = \sum_{i=1}^{n}w_{i}X_{i} =_{d} X_{1}$. Consequently, $VaR_{q}(Z_{{w}}) = VaR_{q}(X_{1})$ is independent of ${w}$ and is the same for all portfolios of risk $X_{i}$ with weights ${w} \in \mathcal{I}_{n}$. 
In other words, in such a case variations in a portfolio has \emph{no effect} on riskiness of its aggregate return. Thus, the symmetric Cauchy distribution with characteristic exponent $\alpha = 1$ is the boundary between extremely heavy-tailed distributions (for which aggregate coverage is statistically not incentive compatible) with infinite first moments, and moderately heavy tailed distributions with finite first moments (aggregate coverage might be sustainable). Similarly, for general weights ${w} = (w_{1},....,w_{n}) \in {\mathbf{R}}^{n}_{+}$, $\alpha = 1$ implies $Z_{{w}} = \sum_{i=1}^{n}w_{i}X_{i} = _{d} (\sum_{i=1}^{n}w_{i})X_{1}$. Thus, $VaR_{q}(Z_{{w}}) = (\sum_{i=1}^{n}w_{i})VaR_{q}(X_{1})$ is independent of ${w}$ so long as $\sum_{i=1}^{n}w_{i}$ is fixed.  Consequently, $VaR_{q}(Z_{{w}})$ is both Schur-convex and Schur-concave in ${w} \in {\mathbf{R}}^{n}_{+}$ for IID $X_{i} \sim S_{1}(\sigma, 0, 0)$. 

 

\subsection{Aggregating IID non-catastrophic cyber-risks}

We now consider aggregating IID risks $X_i$ from the family $\mathcal{\overline{CSLC}}$, which are class of distributions that are convolutions of symmetric distributions that are either log-concave or stable with exponent $\alpha > 1$ - those exhibiting \emph{finite} mean and variance, and representing non-catastrophic heavy-tailed cyber-risks (see \href{https://drive.google.com/file/d/190WaqylxZCTpMCchynYaJ_I17zrBIMfr/view?usp=sharing}{ \underbar{online Appendix A}}   for details). We have the next result regarding VaR performance post cyber-risk aggregation, the proof of which is in \href{https://drive.google.com/file/d/1xHrLRwVOFtP8mWbwf_SuVm-lBmcRPlZO/view?usp=sharing}{ \underbar{online Appendix B}}  . 

\begin{theorem}
\emph{Consider IID r.v's $X_{i} \sim \mathcal{\overline{CSLC}}, i = 1, \cdots, n$, $q \in (0, 1)$, and $n$-vector of weights ${w}, {v} \in {\mathbf{R}}^{n}_{+}$.} 
\emph{Then}
\begin{enumerate}
\item $VaR_{q}(Z_{{w}}) < VaR_{q}(Z_{{v}})$ \emph{if ${v} \prec {w}$ and ${v}$ is not a permutation of ${w}$; in other words, the function $VaR_{q}(Z_{{w}})$ is strictly Schur-convex in ${w} \in {\mathbf{R}}^{n}_{+}$.}

\item \emph{In particular}, $VaR_{q}(Z_{\underbar{w}}) < VaR_{q}(Z_{w}) < VaR_{q}(Z_{\overline{w}})$, $\forall {w} \in \mathcal{I}_{n}$ \emph{such that ${w} \ne \underbar{w}$ and ${w}$ is not a permutation of $\bar{w}$.}
\end{enumerate}
\end{theorem}

\textbf{Theorem Implications} - On a practical note, the theorem simply implies that when an aggregate cyber-risk covering agency is faced with covering independent and identical non-catastrophic cyber-risk distributions, the variance of the combined distribution does not increase with the number of piled up cyber-risks - \emph{simply an encouraging signal for-profit cyber-risk managers to contribute to a sustainable aggregate loss coverage market}. While this latter point has long been believed and empirically validated in the cyber-insurance research literature, the result from Theorem 2.1 is a surprising new facet that we unravel in this paper via theory.

\section{Aggregating Curtailed IID Catastrophic Risks} 
 
In this section we analyze what happens when aggregating multiple heavy-tailed risks each of which has been curtailed, \emph{to fit the realistic scenario where cyber-risk managers have upper bounds on coverage.} We also study the role of how the length of the distributional support needed for the analogue to hold depends on the number of cyber-risks in a manager's portfolio and the degree of heavy-tailedness of unbounded cyber-risk distributions. We have the following result, an analogue of Theorem 2.1 for curtailed catastrophic cyber-risks in this regard, the proof of which is in \href{https://drive.google.com/file/d/1xHrLRwVOFtP8mWbwf_SuVm-lBmcRPlZO/view?usp=sharing}{ \underbar{online Appendix B}}  .  

\begin{theorem}
\emph{Let $n \ge 2$ and let $w\in \mathcal{I}_{n}$ be a weight vector with $w_{[1]} \ne 1$. Let $X_i, i=1, \cdots, n$ be IID r.v.'s $\sim \mathcal{\underline{CS}}(r)$ for some $r\in (0,1)$ and their respective $a$-truncated version given by $Y_i$ defined above. Denote $G(w,z) = P(w_{[1]}X_{1} + w_{[2]}X_{2} > z) - P(X_{1} > z)$, which is positive if $w_{[1]} \ne 1$ (via Theorem 2.1). For any $z > 0$, and all} 
\begin{equation}
\label{eqn:eq31}
    a > \left(\frac{\mathbb{E}[|X_{1}|^{r}](n-1)}{2G(w,z)}\right)^{\frac{1}{r}}, 
\end{equation}
\emph{the following inequality holds:}
\begin{equation}
    P(Y_{w}(a) > z) > P(Y_{1}(a) > z). 
\end{equation}
\emph{Note that $G(w,z)$ reflects that $VaR_{q}[X_{w}] > VaR_{q}[w_{[1]}X_{1} + w_{[2]}X_{2}] > VaR_{q}[X_{1}]$.}
\end{theorem}

The \textbf{implications of this theorem} are multifarious and are presented in multiple blocks. 

\textbf{Implication 1} - The practical implications of the theorem are analogous to Theorem 2.1 in the case of bounded cyber-risks. More specifically, cyber-risk aggregation coverage continues to be disadvantageous in general for catastrophic truncated heavy-tailed distributions.
For $n \ge 2$ and any cyber-risk valuation $z > 0$, there exists $n$ cyber-risks with finite support with the property that the variance return of the aggregate cyber-risk portfolio is riskier than that of the portfolio consisting of a single cyber-risk. From a mathematical viewpoint, Theorems 2.1 and 3.1 indicate that VaR is not sub-additive and, thus, its coherency (see {\href{https://drive.google.com/file/d/190WaqylxZCTpMCchynYaJ_I17zrBIMfr/view?usp=sharing}{ \underbar{online Appendix A}}} for details) is always violated in the class of extremely heavy-tailed cyber-risks with infinite first moments. More specifically, Theorem 3.1 implies that VaR may also be non-coherent in the world of cyber-risks with bounded distributional support. We just proposed conditions under which it is statistically incentive compatible for a (re)-insurer to spread catastrophic cyber-risks having heavy tails. One could also further study conditions under which it will \emph{not be optimal} to spread risks - in the interest of space, this analysis is provided in \href{https://drive.google.com/file/d/14CdCGsX8P2iRJDmcTxidGaqmFOGugdi2/view?usp=sharing}{\underbar{online Appendix C}} and also in \cite{palwsc20}.

\textbf{Implication 2} -  We note that in \emph{the special case of a cyber-risk portfolio with equal weights}, $\tilde{w}_{n} = \left(\frac{1}{n}, \frac{1}{n},....,\frac{1}{n}\right)$, we have
\begin{equation}
\label{eqn:eq40}
    G(\tilde{w}_{n},z) = P\left(\frac{X_{1} + X_{2}}{2} > z\right) - P(X_{1} > z).
\end{equation}
This means that the length of the distributional support reflecting statistical incentive non-compatibility to aggregate cyber-risk coverage in Theorem 3.1 can be taken to be same for all the portfolios with equal weights $\tilde{w}_{n}$. This holds, obviously, for the whole class of the portfolios $w$ such that $w_{[1]} < \frac{1}{2}$. Furthermore, a similar result holds as well for the class of portfolios $w$ such that $w_{[1]} < 1 - \epsilon$, (and, thus, $w_{i} < 1 \epsilon$ for all $i$), where $0 < \epsilon < \frac{1}{2}.$  As follows from the proof of Theorem 3.1, for all such portfolios $w$, the theorem holds for $a > \left(\frac{\mathbb{E}[|X_{1}|^{r}](n-1)}{2\tilde{G}(w,z)}\right)^{\frac{1}{r}}$, where $\tilde{G}(\epsilon, z) = P\left((1 - \epsilon)X_{1} + \epsilon X_{2} > z)\right) < G(w,z)$. This follows since any vector $w$ with $w_{[1]} < 1 - \epsilon$ is majorized (see basics of majorization in the {\href{https://drive.google.com/file/d/190WaqylxZCTpMCchynYaJ_I17zrBIMfr/view?usp=sharing}{ \underbar{online Appendix A}} }) by the vector $(1 - \epsilon, \epsilon, 0,..,0)$. 

\textbf{Implication 3} - From the proof of Theorem 3.1, it follows that, in the special case of portfolios with equal weights $\tilde{w}_{n}=\left(\frac{1}{n}, \frac{1}{n}, \ldots, \frac{1}{n}\right)$ where $n>2,$ the length of the interval of truncation $a$ can be reduced to a smaller value. In such a case, the theorem holds under the restriction $a>\left(\frac{E\left|X_{1}\right|^{r}(n-1)}{2 F_{n}(z)}\right)^{1 / r}$, where
\begin{equation}
F_{n}(z)=P\left(\frac{\sum_{i=1}^{n} X_{i}}{n}>z\right)-P\left(X_{1}>z\right)
\end{equation}
Note that, by Theorem 2.1, $F_{n}(z)>H(z)=G\left(\tilde{w}_{n}, z\right)$ for $n\ge3$.
\emph{This suggests that if the support is large compared to the number of cyber-risks to be aggregated, it might be infeasible for an aggregate risk manager to cover the risks. This demonstrates the ``unpleasant'' properties of VaR as a cyber-risk measure under heavy-tailedness does not arise from the relatively high likelihood of getting very large losses but rather from the fact that there are too few cyber-risks available for the profitable aggregate cyber-risk coverage to work.} 


\textbf{Implication 4} - Theorem 3.1 also shows that, for a specific loss probability $q$, there exists a sufficiently large $a$ such that the value at risk $VaR_{q}[Y_{w}(a)]$ of the return $Y_{w}(a)$ at level $q$ is greater than the value at risk $VaR_{q}[Y_{1}(a)]$ of the return $Y_{1}(a)$ at the same level: $VaR_{q}[Y_{w}(a)] > VaR_{q}[Y_{1}(a)]$. \emph{This highlights the dampening factor to the sustainability of covering aggregate heavy-tailed cyber-risks.}
One should emphasize that the last inequality between the returns $Y_{w}(a)$ and $Y_{1}(a)$ holds for the particular fixed loss probability $q$ and, in the comparisons of the values at risks $VaR_{q}\left[Y_{w}(a)\right]$ and $VaR_{q}\left[Y_{1}(a)\right],$ the length of the interval needed for the reversals of the stylized facts on the portfolio variation depends on $q$ (similar to the fact that in Theorem 3.1, the length of the distributional support $a$ depends on the value of the disaster level $z$ - denoting the degree of heavy-tailedness). This is the crucial qualitative difference of the results in Theorem 3.1 for bounded/curtailed cyber-risk distributions and their implications for the value at risk, from those given by Theorem 2.1 and Theorem 3.1 for unbounded risks, where the inequalities hold for all $z>0$ and all $q \in (0,1)$. 

\textbf{Implication 5} \textbf{(Case of non-identical distributions)} - The analogues of Theorem 2.1 hold for i.i.d. risks $X_{1}, \ldots, X_{n}$ that have skewed extremely thick-tailed stable distributions with infinite first moments: $X_{i} \sim S_{0 < \alpha < 1}(\sigma, \beta, 0), \alpha \in(0,1),$
$\sigma>0, \beta \in[-1,1], i=1, \ldots, n .$ As follows from the proof of Theorem 3.1 (see \href{https://drive.google.com/file/d/1xHrLRwVOFtP8mWbwf_SuVm-lBmcRPlZO/view?usp=sharing}{ \underbar{online Appendix B}}  ), this implies that complete analogues of the results in the present section for bounded versions of symmetric risks from the classes $\mathcal{\underline{CS}}(r)$ continue to hold for truncated extremely heavy-tailed stable distributions $S_{\alpha}(\sigma, \beta, 0)$ with $\alpha \in(0,1), \sigma>0,$ and an arbitrary skewness
parameter $\beta \in[-1,1] .$ In particular, Theorem 3.1 continues to hold for arbitrary skewed risks $X_{i} \sim S_{\alpha}(\sigma, \beta, 0),$ $\alpha \in(0,1), \sigma>0, \beta \in[-1,1]$ if $a>\left(\frac{E\left|X_{1}\right|^{r}(n-1)}{G(w, z)}\right)^{1 / r}$.
\begin{figure}[http]
\centering
 \includegraphics[scale = 0.18, angle=0]{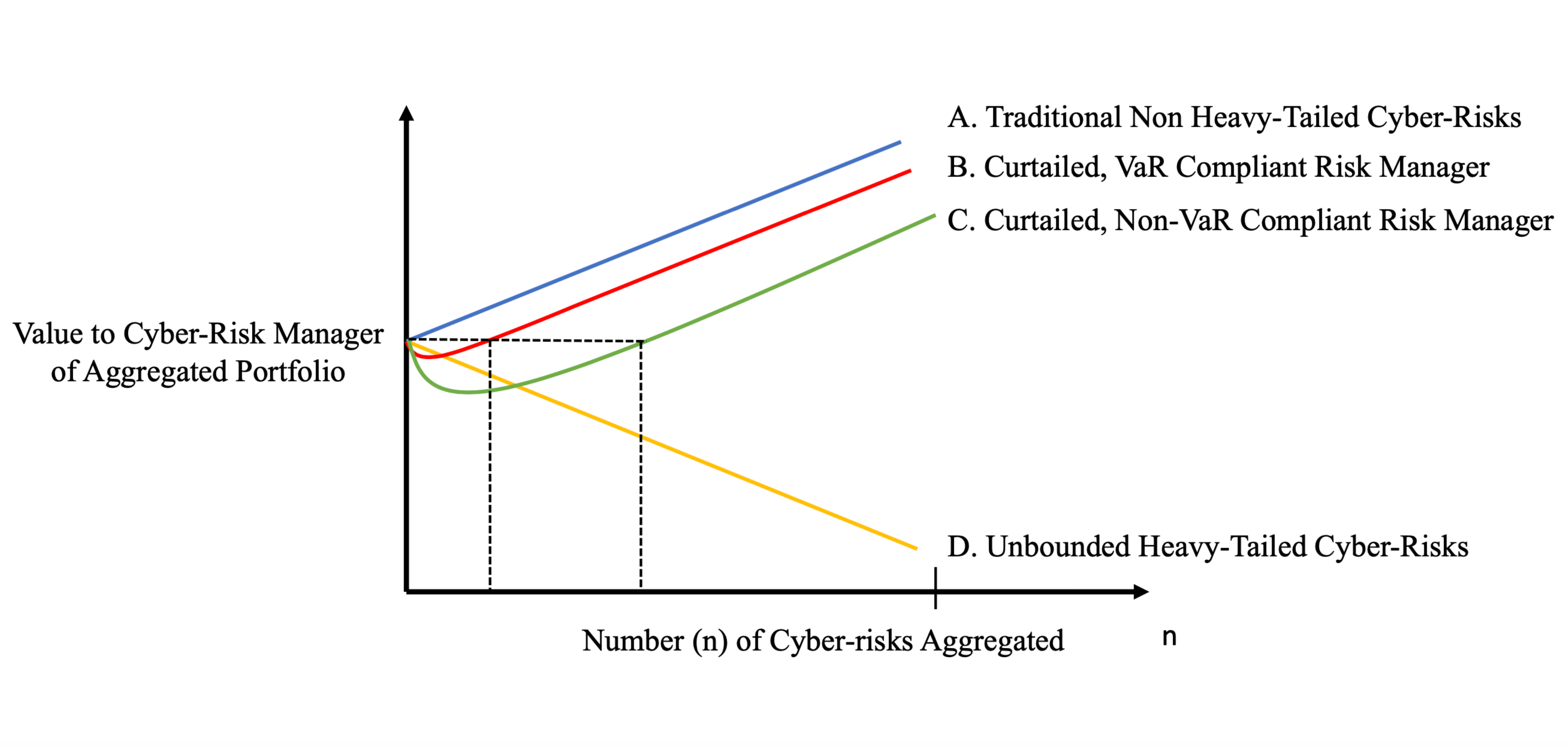}
 \centering
\caption{Conceptual illustration of statistical utility of cyber-risk aggregation as a function of the number of \textbf{i.i.d.} cyber-risks. For a given `n', cyber-risk valuation 'z', there always exists a `dipping' statistical utility region for cases B and C, and the region expands as $a$ increases (ref. Theorem 3.1). $\alpha < 1$ for B, C, and D.}
  \label{fig:1}
\end{figure}

\noindent \textbf{Results Overview and Impact on IoT Societies} - As a summary of the theory results in this section and the previous one, Figure 1 provides a graphical illustration of the impact of the type and number of cyber-risks on a risk manager's valuation (statistical utility, i.e., decreased VaR) of covering aggregate cyber-risk. The interesting observation is that for cases B and C illustrating curtailed cyber-risks, there is a drop in the utility, i.e., increased VaR, as a function of the number of cyber-risks, in covering aggregate risk, followed by an indefinite increase in utility henceforth. The initial drop is due to the tradeoffs from the higher costs of aggregate and increased variance-induced coverage due to a certain threshold `n' catastrophic cyber-risks versus the benefit received from coverage premiums. Clearly, beyond `n' cyber-risks the statistical benefits of aggregate cyber-risk coverage outweighs the negatives of increased risk spread. The outcome of cases A and D are intuitively obvious. 

\noindent In \cite{kessler2014re}, the author rationalizes why aggregate loss coverage services like re-insurance might be sustainable, and not encounter a systemic catastrophe problem. For the general re-insurance setting, he mentions (i) a portfolio of independent risks and geographical diversification, (ii) partial cessation of risk with proper risk screening, and (iii) lack of liability loops, to be the major factors in favor of re-insurance services being sustainable. \emph{However, there are major differences between general re-insurance and cyber re-insurance services, that allows us to closely look at cyber re-insurance service sustainability under universal risk types.} Clearly (i) and (ii) are impractical when major cyber-catastrophes occur and impact IoT societies (e.g., ones caused by the WannaCry and Mirai attacks) (Curve D). In the most optimistic scenarios, Figure 1 illustrates what the size and nature of the coverage portfolio should look like for a cyber re-insurer assuming \emph{limited} coverage liability for i.i.d. heavy-tailed cyber-risks (Curves A-C). However, the challenge still remains to deal with non i.i.d. heavy-tailed cyber-risks such as those posed by \emph{WannaCry} and \emph{Mirai}. 

\section{Aggregating Non-IID Heavy-Tailed Risks}
Cyber-risks are not only heavy tailed in nature, but are likely to be correlated, i.e., tail-dependent. This is true especially in scenarios of major systemic impact causing cyber-attacks. The likelihood of systemic loss impacts are fairly high in a service-networked smart society \cite{palcyber}\cite{coburn2018solving} driven by IoT technologies. In this section we study the effect on VaR on aggregating such cyber-risk types. Statistical correlations and dependencies between distributions are often captured systematically using \emph{copulas} \cite{clemen1999correlations}\cite{danaher2011modeling} (see {\href{https://drive.google.com/file/d/190WaqylxZCTpMCchynYaJ_I17zrBIMfr/view?usp=sharing}{\underbar{online Appendix A}}  } for a preliminary introduction), that are multivariate functions of marginal distributions outputting dependence values. In our case, the marginal distributions are cyber-risk random variables having a heavy-tail characterized via a power-law distribution family.  

To illustrate dependencies between such marginal distributions, we start with the bivariate (generalization to follow) \emph{Eyraud-Farlie-Gumbel-Morgenstern} (EFGM) copula - a power type copula (see {\href{https://drive.google.com/file/d/190WaqylxZCTpMCchynYaJ_I17zrBIMfr/view?usp=sharing}{ \underbar{online Appendix A}}  } for more details) whose marginal distributions obey the power law to reflect heavy-tailed cyber-risk distributions (both catastrophic and otherwise). Let $\left(X_{1}, X_{2}\right)$ be random variables with the EFGM copula and power-law marginals. Then, for any $x \geq 1$ and for $j=1,2$, we have
\[
F_{j}(x)  \sim 1-x^{-\alpha};\,f_{j}(x)  \sim \alpha x^{-\alpha-1}\\
\]
\[
\begin{aligned}
H\left(x_{1}, x_{2}\right) &=\Pi_{i=1}^{2}F_{i}(x_{i})\left[1+\gamma\left(1-F_{1}\left(x_{1}\right)\right)\left(1-F_{2}\left(x_{2}\right)\right)\right] \\
h\left(x_{1}, x_{2}\right) &=\Pi_{i=1}^{2}f_{i}(x_{i})\left[1+\gamma\left(1-2 F_{1}\left(x_{1}\right)\right)\left(1-2 F_{2}\left(x_{2}\right)\right)\right]
\end{aligned}
\]
Let $\left(\xi_{1}(\alpha), \xi_{2}(\alpha)\right)$ be independent random variables from power-law distributions with
tail index $\alpha,$ often called independent copies of $\left(X_{1}, X_{2}\right).$ Our key insight is that in the tail, the behavior of products and powers of power-law densities and distributions of $X_{j}$ 's is identical to the
behavior of their independent copies. This makes it possible to provide asymptotic (with respect to the loss comparisons between the VaR of the aggregated loss and that of a single risk. More
specifically, the crucial component of $\mathbb{P}\left(\frac{X_{1}+X_{2}}{2}>x\right)$ under the EFGM copula can be written as
follows
\[
\begin{aligned}
& \int_{\frac{s+t}{2}>x} \alpha^{2} s^{-\alpha-1} t^{-\alpha-1}\left(2 s^{-\alpha}-1\right)\left(2 t^{-\alpha}-1\right) d s d t \\
=& 4 \alpha^{2} \mathbb{P}\left(\frac{\xi_{1}(2 \alpha)+\xi_{2}(2 \alpha)}{2}>z\right)-2 \alpha^{2} \mathbb{P}\left(\frac{\xi_{1}(2 \alpha)+\xi_{2}(\alpha)}{2}>z\right) \\
&-2 \alpha^{2} \mathbb{P}\left(\frac{\xi_{1}(\alpha)+\xi_{2}(2 \alpha)}{2}>z\right)+\alpha^{2} \mathbb{P}\left(\frac{\xi_{1}(\alpha)+\xi_{2}(\alpha)}{2}>z\right)
\end{aligned}
\]
where the behavior of the individual summands for large $z$ is driven by the lowest tail index of $\xi_{j}$
in the spreading portfolio.

We formalize this result in the following theorem (see \href{https://drive.google.com/file/d/1xHrLRwVOFtP8mWbwf_SuVm-lBmcRPlZO/view?usp=sharing}{ \underbar{online Appendix B}}   for a proof), which generalizes to $n$ dependent heavy-tailed random variables $X_{1}, X_{2}, \ldots, X_{n}$ with multivariate EFGM copula and power-law
marginals.
\begin{theorem}
\emph{For an asymptotically large $z>0,$ and any $n, \alpha>0$}
\[
\mathbb{P}\left(\sum_{i=1}^{n} X_{i}>z n\right) \sim \mathbb{P}\left(\sum_{i=1}^{n} \xi_{i}(\alpha)>z n\right)\]
\end{theorem}
\noindent \textbf{Theorem Implications} - The result suggests that suboptimality of cyber-risk aggregation in the VaR framework for extremely heavy tailed losses carries over from independence to the dependence-capturing EFGM copula. \emph{That is, cyber-risk aggregation increases VaR of dependent extremely heavy tailed risks within this copula family.} It is also easy to see that for dependent losses with the EFGM copula and sufficiently small loss probability $q$, we have
\[
\begin{array}{l}
V a R_{q}\left(\frac{X_{1}+X_{2}}{2}\right)<V a R_{q}\left(X_{1}\right), \quad \text { if } \quad \alpha>1 \\
V a R_{q}\left(\frac{X_{1}+X_{2}}{2}\right)>\operatorname{VaR}_{q}\left(X_{1}\right), \quad \text { if } \quad \alpha<1
\end{array}
\]
Important generalizations of Theorem 4.1 arise if we consider the wider class of power-type copulas. Most popular members of this class such as the polynomial copula of Drouet Mari and Kotz \cite{mari2001correlation} and the copula with cubic section of Nelsen et al. \cite{nelsen1997bivariate} can be written in the following general form
\begin{equation}
C\left(u_{1}, \ldots, u_{n}\right)=\sum_{i_{1}, \ldots, i_{n}=0,1, \ldots} \quad \gamma_{i_{1}, i_{2}, \ldots, i_{n}} \cdot u_{1}^{i_{1}} \cdot u_{2}^{i_{2}} \cdot \ldots \cdot u_{n}^{i_{n}}
\end{equation}
for a multiple index $i=\left(i_{1}, i_{2}, \ldots, i_{n}\right)$ and a set of corresponding parameters $\gamma_{i}$ with appropriate restrictions that make $C\left(u_{1}, \ldots, u_{n}\right)$ a copula. For example, Drouet Mari and Kotz \cite{mari2001correlation}\cite{ibragimov2015heavy} show how to obtain a polynomial copula from function $f=u^{k} v^{q}$. The key feature of such copulas is that they and their densities can be expressed as powers of $u_{j}$ 's. This allows to apply similar arguments as for EFGM. To this end, we have the following theorem, the proof of which is in \href{https://drive.google.com/file/d/1xHrLRwVOFtP8mWbwf_SuVm-lBmcRPlZO/view?usp=sharing}{ \underbar{online Appendix B}}  . 
\begin{theorem}
\emph{For dependent losses with a power-type copula in (5) and for an asymptotically large
$z>0,$ and any $n, \alpha>0,$ the conclusions of Theorem 4.1 hold.}
\end{theorem}
\noindent \textbf{Theorem Implication} - The implications are the same as that of Theorem 4.1.

\section{Experimental Evaluation}
In this section, we put our theory to a rigorous test using real-world cyber-loss data. We want to study whether aggregating individual cyber-risks from different IoT-driven organizational sources (assumed to show characteristics of real-world cyber-loss) in a smart society increase or decrease a risk manager's VaR/Expected Utility (EU) - the scalar metric for measuring the extent of aggregate cyber-risk. 
In a nutshell, we first show using real world data that individual cyber-losses can indeed exhibit a heavy-tailed statistical nature. We then investigate the VaR/EU trends with increasing number of heavy-tailed cyber-risks to be aggregated. 

\subsection{Experimental Setting}
We consider 9015 cyber losses extracted from the publicly available \emph{Privacy Rights Clearinghouse} database, published in 2017. 
We first perform several goodness-of-fit tests for several widely used distributions to characterize the true nature of the cyber-loss distribution. Namely, we use the \emph{normal}, \emph{log-normal}, and \emph{general Pareto} distributions for the purpose of comparison, as in \cite{eling2020extreme}. Based on the goodness-of-fit-statistics (using Log-Likelihood, AIC, BIC, Kolmogorov-Smirnoff, and Anderson-Darling tests), we find that the generalized Pareto distribution fits the data best - thus, . The estimated \emph{Pareto Index} (the exponent in a power law distribution) characterizing a heavy-tailed distribution for the generalized Pareto distribution is 0.1862, using analysis adopted from \cite{nevslehova2006infinite}. 

If a cyber-risk manager (e.g., an insurer) takes on a random risk $X$, a function of $n$ - the number of cyber-risks it accepts to aggregate, the effective outcome (before opting for cyber re-insurance services) for the insurer once $X$ is realized is:
\begin{equation}
\label{eqn:eq1}
    V(x)=\left\{
                \begin{array}{ll}
                 X\, \mathrm{if}\, X < k,\\
                  k\,\mathrm{if}\, X \ge k,
                \end{array}
              \right.
\end{equation}
where $k$ is the limit of the amount of cyber-risk it can accept - true of practice. In the special case when there is no limited liability, i.e., when $k = \infty$, we have $V(X) = X$ for all $X$. If $k < \infty$, $u$ is defined only on $[0, k]$, and without loss of generality $u(k) = 0$. Here, we assume the utility function of a \textbf{perfectly rational} and risk-averse cyber-insurer to be generally of the following form:
\[u(x) = (V(x))^{\beta},\,\beta \in (0,1),\]
which is the power utility function, and for $x$ being a risk variable, is a Von-Neumann Morgenstern (VNM) utility function. $\beta$ is degree of risk-aversion of the cyber-insurer. We perform 100,000 Monte Carlo simulations to obtain our results. 

\subsection{Experimental Results}
\begin{figure}
  \centering\includegraphics[width=1\linewidth]{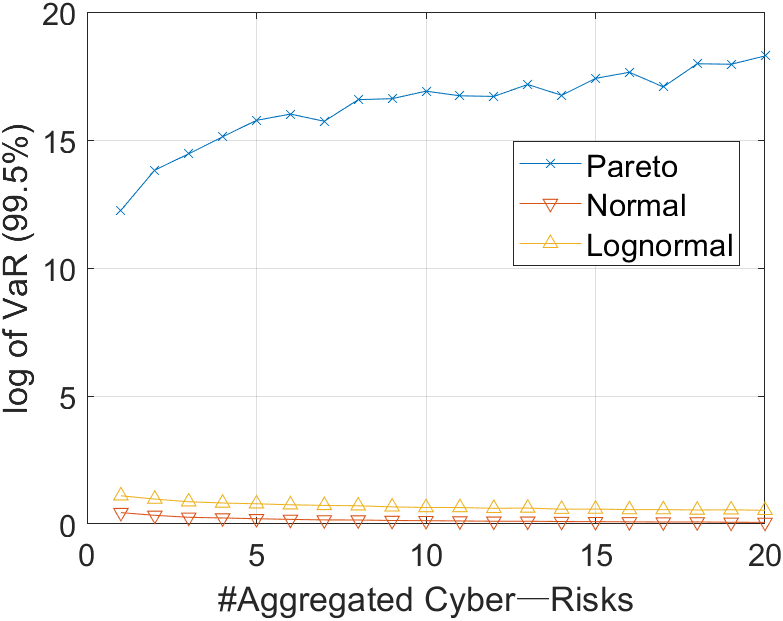}
  \caption*{\textbf{Fig. 2: } Cyber-Risk Aggregation Performance (on the VaR Metric) for \textbf{i.i.d.} Risks of Pareto, Log-Normal, and Normal Distributions.}
  \label{fig:fig1}
\end{figure}

We observe from Figure 2 that $VaR_{0.995}(X)$ monotonically decreases for normal and log-normal individual cyber-risk distributions (fitted using our data set) - though the VaR for log-normal risks decreases, at a slower rate. On the other hand, $VaR_{0.995}(X)$ (denoted as VaR from now on throughout the section) increases (not monotonically) for Pareto individual cyber-risk distributions, as is expected from theory (for symmetric stable cyber-risk distributions). \emph{However, the non-monotonicity indicates (also in accordance to our theory) that for heavy-tailed cyber-risks simulated in practice, there exists a certain number of sampled risks, aggregating which does not increase VaR.} To focus on our empirical data set, we use statistical bootstrapping to simulate the VaR for varying number of aggregated cyber-risks. In this regard, we draw directly from our original sample instead of the different distributions assumed above. The sample is drawn with replacement (thus, i.i.d.) and is of equal size as the original data set (m=9015 observations).
Due to the symmetric stable nature of the cyber-risk distribution induced by the empirical dataset, the Conditional-Value-of-Risk (CVaR) measure provides similar performance (see Figure 3a.) as the VaR measure - as the VaR measure is a coherent risk measure for symmetric and stable risk distributions \cite{mcneil2015quantitative}. 
Moreover, we calculate the confidence interval by repeating the bootstrapping itself.
Figure 3b. shows the bootstrapped VaR and its confidence interval. We observe that the bootstrapped VaR (induced by the empirical loss distribution) always lies above the log-normal VaR and the aggregation benefit is much less prevalent than assumed. As a consequence, \emph{in accordance with theory, not to aggregate heavy-tailed risks at all would be optimal from a cyber-risk management perspective.}  
\par \emph{We now focus on an expected utility (EU) setting} induced on limited liability where applicable, to assess cyber-risk aggregation performance. Figures 4a and 4b show the EU-theoretic performance based on a power utility function $u(x)$ for aggregating i.i.d. cyber-risks. 
As expected, for log-concave cyber-risk distributions, i.e.., for normally distributed and log-normal i.i.d. cyber-risks (Figure 4a), we do not observe  a change in the derivative of the expected utility with increase in the number of cyber-risks aggregated. However, this is not true for a heavy tailed distribution such as the one induced by the empirical dataset and fitted to a Pareto distribution (see Figure 4b.). More specifically, the rate of decrease (corroborated via theory) in expected utility with heavy-tailed cyber-risk distributions fluctuates (instead of exhibiting monotonic behavior, as is usual in sampling scenarios) - however, on average is borderline negative with a high standard deviation.
\par We also study the role of pool of homogeneous cyber-risk managers (CRMs) that share\footnote{We do not explicitly consider the strategic aspects of sharing in this work.} aggregate cyber-risk (e.g., like in a cyber re-insurance business), on the EU of a single manager in that pool. 
We consider two instances of individual cyber-risks - one  with a synthetic Pareto index $\alpha$ that is 1 (characterizing heavy-tail nature of cyber-risk), and one lying below 1 (characterizing extremely heavy-tailed cyber-risks), that is characterized by our real-world data set.
Figure 5 shows that for risk with a Pareto Index of 1 and limited liability of $k$ = 60, the expected utility of a single manager for different aggregation and cyber-risk pooling sizes (\#CRMs), is U-shaped. The U-shape denotes that the benefit from aggregation first decreases before it eventually increases again (similar trend to that in Figure 1 that illustrates our theory). 
Using a Pareto index of 0.1862 (as estimated from the data, and indicating an extreme heavy-tailed distribution) changes, \emph{ceteris paribus}, the result completely, as shown in Figure 6a. and 6b. \emph{Since the expected utility decreases monotonically (at nearly a constant rate of decrease) not providing any (pooled) coverage management such as insurance would be optimal and the aggregate coverage market would fail completely.} Our numerical analysis shows that the U-shape can only be observed if the Pareto tail index is in the range of (0.8, 1.12)[model-based]. While the situation in Figure 5 leaves room for traditional cyber (re)insurance promoting regulatory intervention that enables curtailed heavy-tailed cyber-risk distributions sourced from organizations to have a tail-index in a feasible range, the one in Figure 6 suggests otherwise for a risk-aggregating cyber re-insurance business that does not curtail very heavy-tailed cyber-risks. \emph{More precisely, cyber-risk pooling is not business-beneficial for cyber-risk managers (CRMs), if individual cyber-risks are heavy-tailed cyber-risks (unless these risks are curtailed), and the subsequent coverage market fails.}

\noindent \textbf{An Important Note on the Results} - The data used in this paper is not generated from an IoT system. The data captures heavy-tailed properties of cyber-loss distributions that we take advantage of to show whether it is feasible to aggregate individual cyber-risks having such properties. There is no real world data available pertaining to IoT systems as of yet, to the best of our knowledge, that reflects heavy-tailed cyber-loss distributions - though in principle it is fair to assume that some IoT-related cyber-loss data sets would exhibit heavy-tailed characteristics.

 \begin{center}
\begin{figure*}[!ht]
\begin{minipage}{.5\textwidth}
\begin{subfigure}{\textwidth}
\centering
  \includegraphics[width=0.95\linewidth]{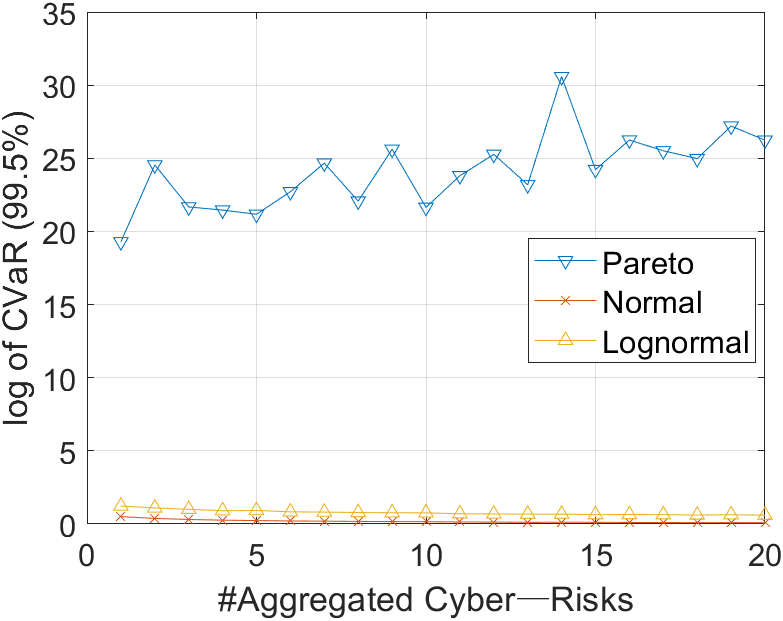}
  \label{fig:example4}
\end{subfigure}
\end{minipage}
\begin{minipage}{.5\textwidth}
\begin{subfigure}{\textwidth}
  \centering
  \includegraphics[width=0.95\linewidth]{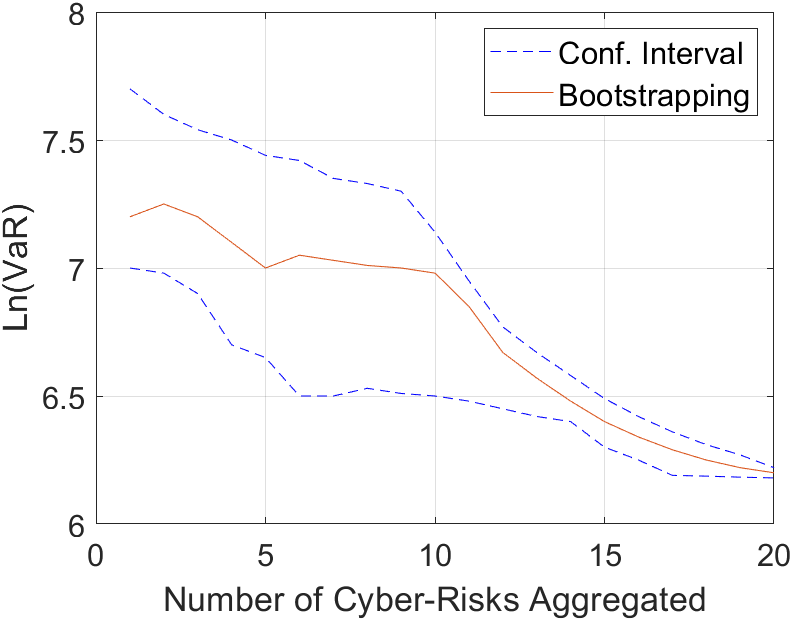}
 \label{fig:example7}
\end{subfigure}
\end{minipage}
\vspace{-2 mm}
  \caption*{\textbf{Fig. 3: }(a) Cyber-Risk Aggregation Performance (on the CVaR Metric) for \textbf{i.i.d.} Risks of Pareto, Log-Normal, and Normal Distributions, (b) Confidence Intervals of Cyber-Risk Aggregation Performance (VaR) for \textbf{i.i.d.} Cyber-Risks Characterized by the Real-World Dataset.}
  \vspace{-3 mm}
\end{figure*}
\vspace{-2 mm}
\end{center}
 \begin{center}
\begin{figure*}[!htbp]
\begin{minipage}{.5\textwidth}
\begin{subfigure}{\textwidth}
  \centering
  \includegraphics[width=0.95\linewidth]{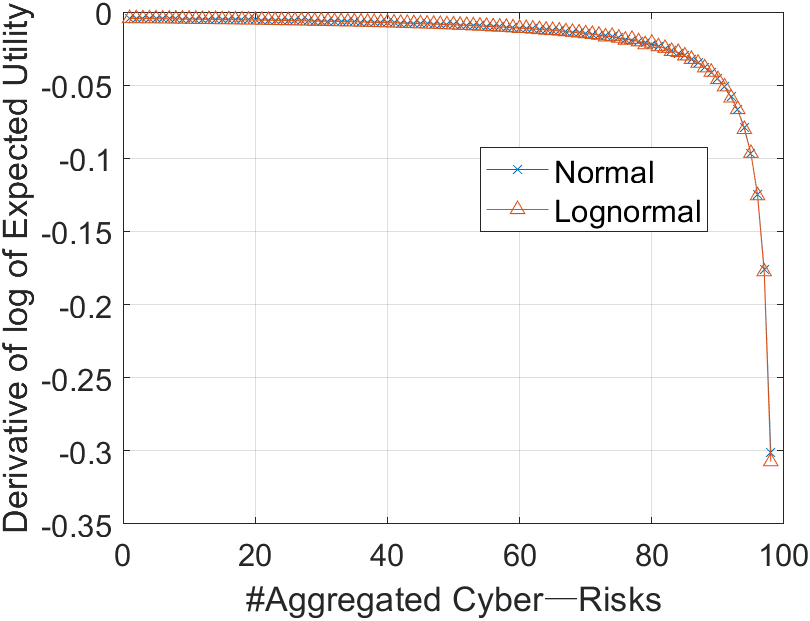}  \vspace{-2 mm}
  \label{fig:example4}
\end{subfigure}
\end{minipage}
\begin{minipage}{.5\textwidth}
\begin{subfigure}{\textwidth}
  \centering
  \includegraphics[width=0.95\linewidth]{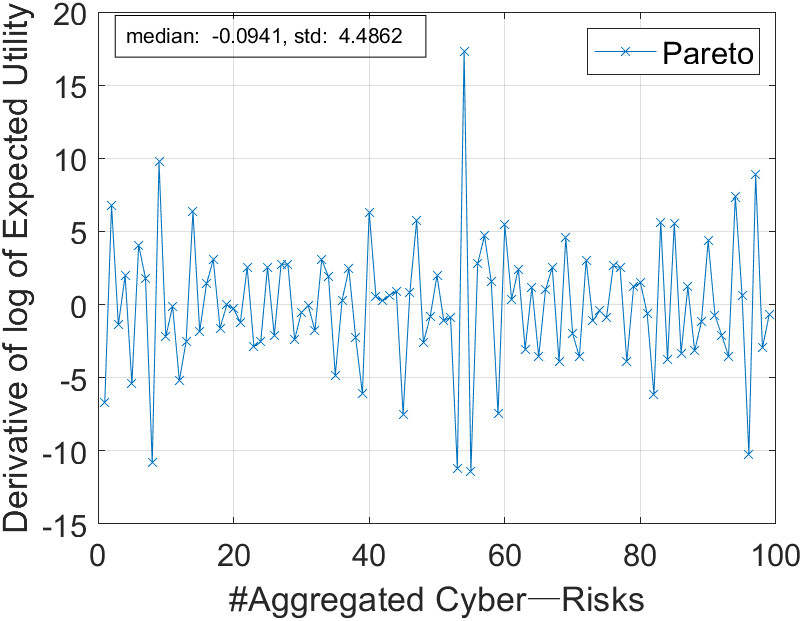}
\vspace{-2 mm}
 \label{fig:example7}
\end{subfigure}
\end{minipage}
  \caption*{\textbf{Fig. 4: }Cyber-Risk Aggregation Performance (on the Expected Utility Metric) for (a) \textbf{i.i.d.} Risks of Different Log-Concave Distributions, and (b) \textbf{i.i.d.} Heavy-Tailed Pareto Distributions.}
\end{figure*}
\vspace{-4 mm}
\end{center}
\vspace{-10 mm}

\section{Related Work}
In this section, we solely focus on research related to cyber-risk aggregation. We partition this section in two parts: (i) the heavy-tailed and tail-dependent nature of cyber-risk, and (ii) feasibility insights regarding the profitable coverage of aggregate heavy-tailed cyber-risk. The readers are referred to \cite{romanosky2019content}\cite{xu2019cybersecurity} for references to research on pricing cyber-risk. 

\vspace{-2 mm}
\subsection{On the Heavy-Tailed and Dependent Nature of Cyber-Risk}
There are quite a few instances in practice where cyber-risks have shown heavy-tailed impact. In \cite{maillart2010heavy}, Maillart and Sornette analyzed a \emph{Datalossdb} 2017 dataset consisting of 956 personal identity loss incidents that occurred in the United States between year 2000 and 2008. They found that the personal identity losses per incident, denoted by $X$, can be modeled by a heavy tail
distribution $P(X > n) \sim n^{-\alpha}$ where $\alpha$ = 0.7 +/- 0.1, 
and more importantly this result holds for a variety of organizations: business, education, government, or a medical institution. Because the probability density function of the identity losses per incident is static, the situation of identity loss is stable from the point of view of the breach size.  Edwards et al. \cite{edwards2016hype} analyzed a Privacy Rights Clearinghouse database of 2017 consisting of 2,253 breach incidents that span over a decade from 2005 to 2015. These breach incidents include two categories: negligent breaches (i.e., incidents caused by lost, discarded, stolen devices, or other reasons) and malicious breaching (i.e., incidents caused by hacking, insider and other reasons). They showed that the breach size can be modeled by log-normal or log-skewnormal distribution that are heavy-tailed distributions, and the breach frequency can be modeled by the negative binomial distribution. In \cite{wheatley2016extreme}, Wheatley et.al., merged and analyzed cyber-breach incidents from the Datalossdb and the Privacy Rights Clearinghouse database spanning over a decade (2000 to 2015). They used the Extreme Value Theory (EVT) \cite{embrechts2013modelling} to study the maximum breach size, and further modeled the large breach sizes by a doubly truncated heavy-tailed Pareto distribution. 
There are also studies establishing the dependence among cyber risks. Notable among them are \cite{herath2011copula}\cite{pal2019robust}\cite{mukhopadhyay2013cyber}\cite{bohme2006models}\cite{xu2017vine}\cite{xu2018modeling}\cite{xu2019cybersecurity}\cite{peng2018modeling}. 

\noindent \textbf{Shortcomings} - Existing research in cyber-security has been successful in elucidating the heavy-tailed and tail-dependent nature of cyber-risk; however, \emph{is yet to propose formally proven directions to allow a profit-minded cyber-risk manager to judge whether a collection of such risks is suitable to aggregate, under various degrees of heavy-tailedness.} This decision making problem will increasingly arise in the IoT age where major cyber-risks affecting smart societies will give rise to a systemic effects that cyber-risk managers have to deal with. It is a common perception from empirical studies and insurance literature that i.i.d. cyber-risks, even though heavy-tailed, are suitable for aggregation. In this paper, we showed quite the contrast for i.i.d. catastrophic heavy-tailed risks.

\vspace{-3 mm}
\subsection{Covering Aggregate Cyber-Risk in IoT Societies}
In a recent work, a group of researchers \cite{palcyber} have studied the problem of whether (a) the underlying network of service organizations in society relying on IT/IoT technologies, and (b) the statistical nature of cyber-risk distributions, positively or negatively affect aggregate cyber-risk managers in expanding their business. The authors surprisingly show that both, the underlying network, as well as i.i.d. and non i.i.d. non-heavy tailed cyber-risk distributions does not have a major role to play (does not imply independence) in encouraging or discouraging aggregate cyber-risk managers to expand or contract their coverage business. 

\noindent \textbf{Shortcomings} - The cited work, though tackling the problem of judging the role of the network and the nature of cyber-risk distributions on the future of cyber-risk aggregation business, does not model catastrophic and tail-dependent heavy-tailed cyber-risks that may be a possibility in modern IoT-driven societies. However, as a major positive, their result in the work does provide confidence to aggregate cyber-risk managers to boost their cyber-loss coverage business for non-heavy tailed cyber-risks in a networked interdependent setting - something the digital society is in need of.  

\begin{figure}
  {\centering\includegraphics[width=1\linewidth]{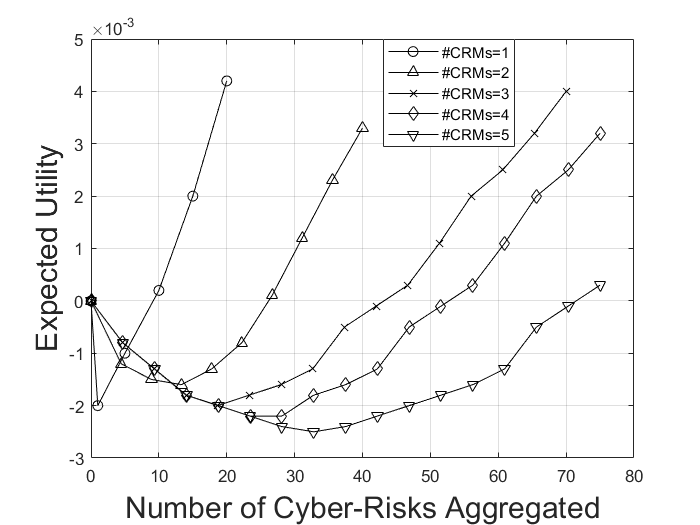}
  \caption*{\textbf{Fig. 5: } Curtailed Cyber-Risk Aggregation Performance (on the Expected Utility Metric) for i.i.d. Pareto Risk Distributions, with varying number of CRMs. Here, $k$ = 70, $\beta$ = 0.0315, $\alpha$ = 1.}
  \label{fig:fig5}}
\end{figure}

\section{Discussion and Summary}
{In this section, we first provide a brief review of the current state of insurance-driven CRM (an indicator of the degree of cyber-risk control) in small and medium IT-driven businesses that represent the majority of IT businesses in operation, and gauge the likelihood of cyber-risk distributions that may be sourced at these businesses. More importantly SMBs are highly service networked among themselves, and this network can pose significant cyber-risk aggregation challenges for CRM solution providers \cite{palcyber}. Our review is based on recent \emph{Advisen} and \emph{CyberScout} reports - indsutry leaders in CRM and cyber-security solutions. Finally, we summarize the paper. 

\begin{center}
\begin{figure*}[ht]
\begin{minipage}{.5\textwidth}
\begin{subfigure}{\textwidth}
  \centering
  \includegraphics[width=0.95\linewidth]{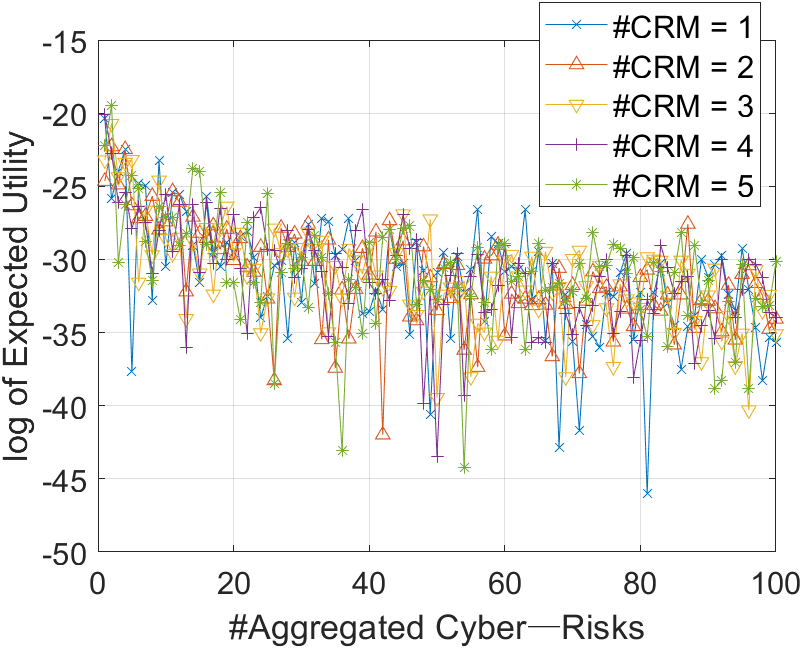}  
  \label{fig:example4}
\end{subfigure}
\end{minipage}
\begin{minipage}{.5\textwidth}
\begin{subfigure}{\textwidth}
  \centering
  \includegraphics[width=0.95\linewidth]{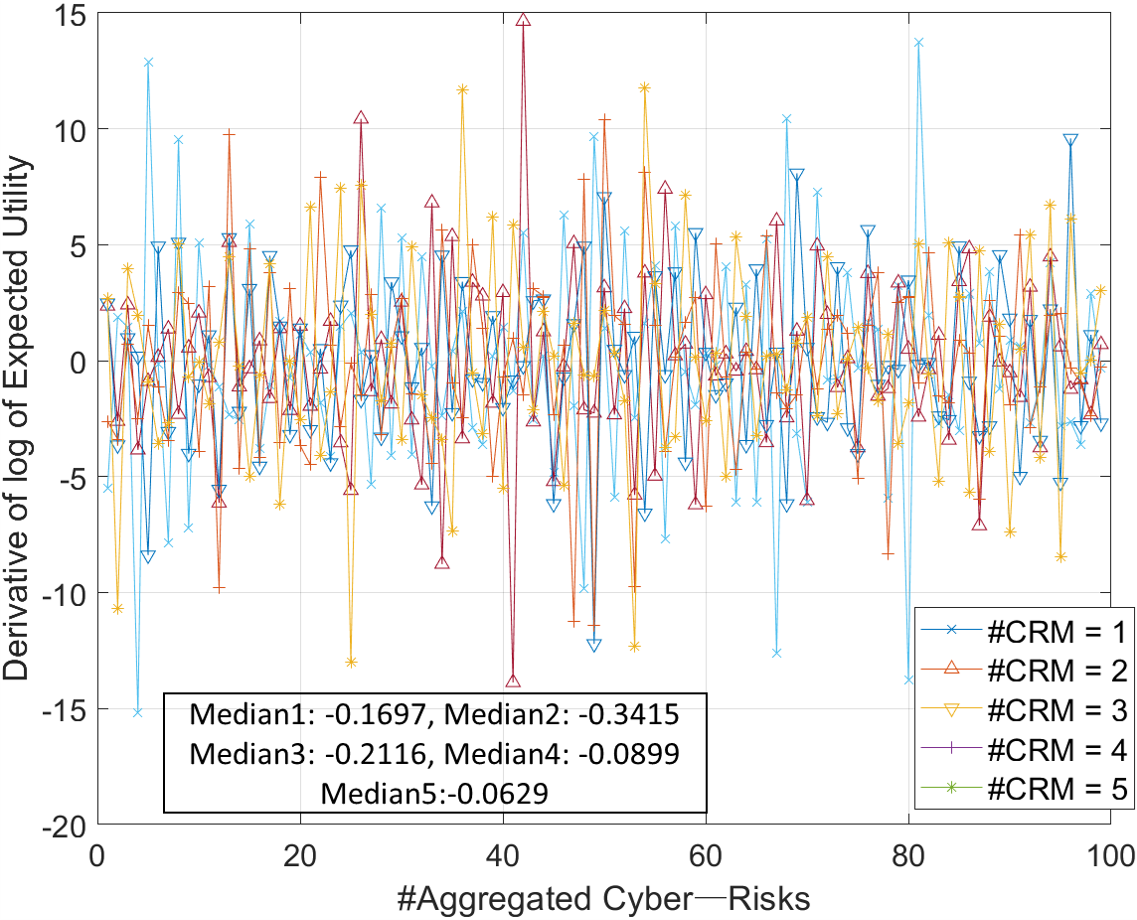}
 \label{fig:example7}
\end{subfigure}
\end{minipage}
  \caption*{\textbf{Fig. 6: }Curtailed Cyber-Risk Aggregation Performance (on the Expected Utility Metric) for i.i.d. Pareto Risk Distributions, with varying number of CRMs. Here, $k$ = 70, $\beta$ = 0.0315,  $\alpha$ = 0.1862.}
\end{figure*}
\end{center}
\vspace{-10 mm}
\subsection{Discussion} 
Small and medium-sized businesses are an important driver of the economy and should be empowered with progressive insurance policies that include cyber risk protection services, incident response and insurance coverages to provide the financial support needed to keep the doors open after an attack. As of 2020, insurers and cybersecurity services firms are innovating around the clock to create risk mitigation policies and procedures that can provide peace of mind to SMB leaders. However, despite a rise in cyberattacks against small and mid-size businesses, about 69\% of SMB respondents to a recent survey by CyberScout said they did not carry cyber insurance coverage and worryingly many don't even have the appropriate security safeguards in place - clearly indicating a lack of seriousness by SMBs to improve their cyber-hygiene. Moreover, in the age of COVID, business owners are under a lot of pressure from the economic disruptions caused by the pandemic, and finding it even more challenging now to find the time to prioritize cyber-security. CyberScout found that 16\% of the respondents had experienced a ransomware event and 40\% said they would not know who to contact if they did fall victim to ransomware. SMBs also may not be aware enough of the ransomware risk – data breach ranks as the highest concern for 30\% of respondents, but ransomware is tops for only 10\%. And only 22\% have a backup plan in place. Over half (51\%) of survey respondents had no formal cyber-security training program, but 76\% said they felt confident about their company’s security infrastructure. However, the results revealed some possible gaps. A quarter of respondents said they send out ``best practices" emails to employees, 22\% reported performing ``live fire" trainings and 20 percent also performed vulnerability testing. Annual trainings were the only measure taken by 18\% of the respondents. Due to the pandemic, just over half (53\%) reported having employees work remotely, but only 34\% required
the use of a VPN connection and only 17\% took any steps to create or remind employees of remote work security protocols. In fact, 14\% said they had no specific cyber measures for remote working. \emph{Clearly, even in 2020, the state of cyber-security strength in SMBs is far from desired, and there is a significant likelihood of each being a source of heavy-tailed, i.e., catastrophic, cyber-risks in the event of major cyber-attacks.}}

\vspace{-2mm}
\subsection{Paper Summary}
In this paper, we provided a rigorous general theory to elicit conditions on (tail-dependent) heavy-tailed cyber-risk distributions under which a risk management firm will find it (un)profitable to provide aggregate cyber-risk coverage for IoT-driven smart societies. As our primary novel contributions, we proved that (a) spreading \emph{catastrophic} heavy-tailed cyber-risks that are identical and independently distributed (i.i.d.), i.e., not tail-dependent, \emph{is not} an effective practice for aggregate cyber-risk managers, whereas spreading \emph{non-catastrophic} i.i.d. heavy-tailed cyber-risks is, and (b) spreading \emph{catastrophic and tail-dependent} heavy-tailed cyber-risks \emph{is not} an effective practice for aggregate cyber-risk managers. \emph{A summary of cyber-risk management effectiveness results for various i.i.d./non-i.i.d. distributions is shown in Figure 7.} We conducted a real-data driven numerical study to validate claims made in theory.
\vspace{-2mm}
\begin{figure}
  {\centering\includegraphics[width=0.8\linewidth]{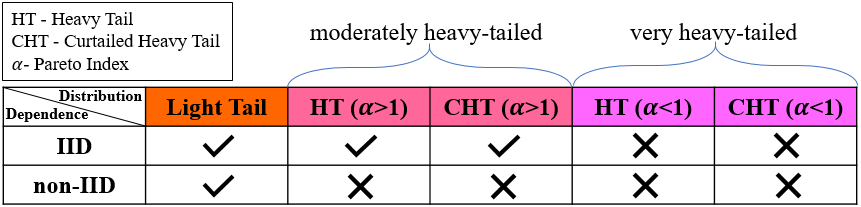}
  \caption*{\textbf{Fig. 7:} Summary of The Effectiveness (Yes (Tick)/No (Cross)) of Aggregate (Large Enough $n$) Cyber-Risk Management for Light and Heavy-Tailed IID/non-IID Distributions.}}
  \label{fig:fig}
\end{figure}

\section*{ACKNOWLEDGMENTS}
This work has been supported by the NSF under grants CNS-1616575, CNS-1939006, and ARO W911NF1810208.

\bibliography{acm-bibfile,cyber-elsevier}
\bibliographystyle{ieeetr}

\vfill

\end{document}